%Paper: hep-th/9307079
%From: andy@nsfitp.itp.ucsb.edu (Andrew Strominger)
%Date: Sun, 11 Jul 93 12:13:20 PDT

%%%%%%%%%%%%%%%%%%%%%%%%%%%%%%%% NOTE %%%%%%%%%%%%%%%%%%%%%%%%%%%%%%%%%%%%%
% This paper has four figures appended as a second part in a uuencoded,     %
% tar compressed postscript file with instructions for unpacking.         %
% These figures must be printed out separately.                             %
%%%%%%%%%%%%%%%%%%%%%%%%%%%%%%%%%%%%%%%%%%%%%%%%%%%%%%%%%%%%%%%%%%%%%%%%%%%

\input jnl.tex

\def\ie{{\it i.e.,\ }}
\def\a{\alpha}

\def\de{{\delta}}
\def\om{{\omega}}

\def\la{\lambda}

\def\om{{\omega}}
\def\Th{{\Theta}}

\def\Om{{\Omega}}

\def\bbR{{I\kern-0.3em R}}

\def\pmb#1{\setbox0=\hbox{$#1$}%
\kern-.025em\copy0\kern-\wd0
\kern.05em\copy0\kern-\wd0
\kern-.025em\raise.0433em\box0 }

\def\q2{{Q^2}}
\def\gtwid{\raise.3ex\hbox{$>$\kern-.75em\lower1ex\hbox{$\sim$}}}
\def\ltwid{\raise.3ex\hbox{$<$\kern-.75em\lower1ex\hbox{$\sim$}}}
\def\12{{1\over2}}

\def\part{\partial}

\def\topppageno1{\global\footline={\hfil}\global\headline
={\ifnum\pageno<\firstpageno{\hfil}\else{\hss\twelverm --\ \folio
\ --\hss}\fi}}

\def\toppageno2{\global\footline={\hfil}\global\headline
={\ifnum\pageno<\firstpageno{\hfil}\else{\rightline{\hfill\hfill
\twelverm \ \folio
\ \hss}}\fi}}

\def\boxit#1{\vbox{\hrule\hbox{\vrule\kern3pt
  \vbox{\kern3pt#1\kern3pt}\kern3pt\vrule}\hrule}}

\def\slS{\raise.15ex\hbox{$/$}\kern-.53em\hbox{$S$}}
\def\cI{{\cal I}}
\def\cH{{\cal H}}
\def\cD{{\cal D}}

\rightline{NSF-ITP-93-92}
\rightline{hep-th/9307079}

\title WHITE HOLES, BLACK HOLES AND CPT IN TWO DIMENSIONS

\author Andrew Strominger

\affil Department of Physics
University of California at Santa Barbara, CA 93106-9530
and
Institute for Theoretical Physics
University of California at Santa Barbara, CA 93106-4030

\abstract{
It is argued that a unitarity-violating but weakly
CPT invariant superscattering
matrix exists for leading-order
large-$N$ dilaton gravity, if and only if one includes in the Hilbert space
planckian ``thunderpop" excitations which create white holes.
CPT apparently cannot be realized in a low-energy effective theory
in which such states have been integrated out. Rules
for computing the leading-large-$N$ superscattering are described in terms of
quantum field theory on a single multiply-connected spacetime  obtained by
sewing the future (past) horizons of the original spacetime with the
past (future) horizons of its CPT
conjugate. Some difficulties which may arise in going
beyond leading order in
$1/N$ are briefly discussed.
}
\endpage
\head{1. Introduction}

It has been forcefully argued [\cite{hawk}] that the laws of quantum mechanics,
when applied to black holes, imply the breakdown of unitarity, at least for
observers outside the black hole who cannot tell whether the
the information has been destroyed at a singularity, carried away by a baby
universe, or inaccessibly stored in a
remnant. The $S$-matrix must then be replaced by a
superscattering matrix  $\slS$ which in
general maps pure states into mixed states
(but conserves probability). Detailed analyses of two-dimensional toy models
over the last year and a half[\cite{CGHS,emod,rst,gine,lowe,pira}]
have greatly fortified the arguments of
[\cite{hawk}] -- although there are some
new wrinkles [\cite{bos}]. At least in some models information is effectively
lost. Thus the role of unitarity as a cornerstone of physics is in serious
jeopardy.

CPT is also considered a sacred principle of physics. However, its validity is
also seriously threatened in the context of quantum black hole processes.
Indeed, it is easy to see [\cite{page}] that unitarity violations imply a
breakdown of {\it strong\/} CPT. Strong CPT states that if a given in density
matrix $\rho^A_{in}$ is mapped to a given out density matrix $\rho^B_{out}$;
$$
\rho^B_{out}=\slS \rho^A_{in}, \eqno(gone)
$$
then the CPT reverse of $\rho^B_{out}$ is mapped to the CPT reverse of
$\rho^A_{out}$:
$$
\Th\rho^A_{in}=\slS\Th^{-1}\rho^B_{out}, \eqno(gtwo)
$$
where $\Th(\Th^{-1})$ is the CPT map from in (out) to out (in) density
matrices. Multiplying both sides of \(gtwo) by $\Th^{-1}$ and inserting it into
the right hand side of \(gone) one learns that
$$
\slS^{-1}=\Th^{-1}\slS \Th^{-1} \eqno(scpt)
$$
is the inverse of $\slS$. However, it is easily seen [\cite{page}], that if
pure states go to mixed states, information is lost and $\slS$ therefore can
not have an inverse.

Put another way, the $\slS$-matrix for black hole processes contains an
intrinsic arrow of  time: it takes pure
states to mixed states but not mixed states to pure states. This is
inconsistent with strong CPT.

However Wald [\cite{wald}] has pointed out that strong CPT violation is really
a consequence of our CPT non-invariant description of the dynamics, rather than
of a fundamental arrow of time. To see this consider [\cite{sork}]
a $\slS$-matrix which
maps any state of a spin $1/2$ particle to either an up or down
state, both with probability one half. Clearly this involves no fundamental
arrow of time: one could not discern whether a movie of the scattering process
was being run forward or backward. Physical discernable violations of CPT
must violate the weak CPT condition. This states that the probability to
scatter from a given pure in-state $\rho^P_{in}$ to a pure out-state
$\rho^Q_{out}$ is equal to the probability of scattering from
$\Th^{-1}\rho^Q_{out}$ to $\Th\rho^P_{in}$:
$$
\rho^Q_{out}\slS\rho^P_{in}=(\Th\rho^P_{in}))\slS(\Th^{-1}\rho^Q_{out}).
\eqno(xwz)
$$
Equivalently,
$$
\slS^T=\Th^{-1}\slS\Th^{-1}. \eqno(wcpt)
$$
Note that \(wcpt) and \(scpt) are different. \(wcpt) is obeyed in the given
example of the spin $1/2$ particle.

Wald has speculated that the $\slS$-matrix for black hole formation and
evaporation obeys the weak CPT condition. An immediate  consequence of this
would be that the CPT reverse of the final state of black hole evaporation
(consisting mostly of outgoing Hawking radiation) usually
collapses to form a black
hole [\cite{wald}].

In this paper we will show that this is not the case in a two dimensional
model. CPT reversed Hawking radiation is below threshold for black hole
formation: it is reflected through the origin. We strongly suspect that this is
also the case in four dimensions.

However we shall see in two dimensions that weak CPT invariance can be
restored---at least in a sector of the Hilbert space---by including the
possibility of white hole formation/evaporation\footnote{*}{This possibility
was rejected by Wald[\cite{wald}] on the
basis of white hole instabilities [\cite{wbld}]. These instabilities are
suppressed in a $1/N$ expansion, but as shall be discussed in
section 5, they may
create difficulties at finite $N$. }. This process has a
semiclassical description which is basically the time reverse of black hole
formation/evaporation . Black holes lead to unpredictability because
information is lost behind the future event horizon, while white holes lead to
unpredictability because one must integrate over initial conditions at the past
event horizon. We will find that the integration measure is fixed by weak CPT
invariance. An intriguing representation of the $\slS$-matrix will be given as
an $S$-matrix on a (in general multiply-connected) spacetime obtained by
sewing together white and black hole horizons of the original
spacetime with its time reverse.

\head{2. Brief Review of the RST Model}

An elegant model for two-dimensional black hole evaporation was
introduced by Russo, Susskind and Thorlacius [\cite{rst}], expanding on ideas
introduced in [\cite{emod}]. The RST model differs from the original
CGHS model[\cite{CGHS}]
by finite counterterms which are fine-tuned to preserve a global symmetry. This
enables an exact solution of the theory at large $N$. Numerical analyses of the
CGHS model [\cite{lowe,pira}] indicate that the models are qualitatively
similar, despite the fine-tuning.

The classical Lagrangian for the RST model is, in conformal gauge,
$$
\eqalign{S_{cl} ={1\over\pi} \int d^2x \Bigg[&(2e^{-2\phi}-{N\over12}\phi)
\part_+\part_-\rho\cr
&+e^{-2\phi}(\la^2e^{2\rho}-4\part_+\phi\part_-\phi)
+{1\over2}\sum^N_{i=1}\part_+f_i\part_-f_i\Bigg],\cr} \eqno(cact)
$$
where $\rho$ is the conformal factor, $\phi$ is the dilaton and $f_i$ are $N$
scalar  matter fields. Define\footnote{*}{Our conventions differ slightly from
[\cite{rst}]. They are chosen so that  $\chi$ and $\Om$ are
held fixed as $N$ is taken to infinity.}
$$
\eqalignno{
\Om &={12\over N}e^{-2\phi}+{\phi\over2}+{1\over4}\ln{N\over 48}, &(odef)\cr
\chi &={12\over N}e^{-2\phi}+\rho-{\phi\over2}-{1\over4}\ln{N\over 3}. &(cdef)
\cr}
$$
In the large-$N$ limit, with $\chi$ and $\Om$ held fixed, the quantum effective
action is then
$$
S ={1\over\pi}\int d^2 x\Bigg[{N\over12}(-\part_-\chi
\part_+\chi+\part_+\Om\part_-\Om
+\la^2e^{2\chi-2\Om})+{1\over2}\sum^N_{i=1}\part_+f_i\part_-f_i\Bigg].
\eqno(nact)
$$
When rewritten in terms of $\rho$ and $\phi$, \(nact) is seen to differ from
the classical action \(cact) by the term ${N\over12}\part_+\rho\part_-\rho$
responsible for
Hawking radiation. (The effects of ghosts may be ignored in the large-$N$
limit.)

There is a residual conformal gauge invariance in \(nact). We fix
this by the "Kruskal gauge" choice
$$
\chi=\Om, \eqno(gchc)
$$
which implies
$$
\rho=\phi+{1\over2}\ln{N\over12}. \eqno(rphi)
$$
In Kruskal gauge the equations of motion are simply
$$
\part_+\part_-\Om=-\la^2, \eqno(oeom)
$$
and the constraints
reduce to
$$
\part^2_\pm\Om=-T^M_{\pm\pm}, \eqno(cstr)
$$
where
$$
T^M_{\pm\pm}={6\over N}\sum^N_{i=1}\part_\pm f_i\part_\pm f_i+t_\pm.
\eqno(tdef)
$$
The functions $t_\pm(x^\pm)$ are fixed by boundary conditions.

The linear dilaton vacuum solution
$$
\phi=-{1\over2}\ln\Bigg[{-\la^2 Nx^+x^-\over12}\Bigg] \eqno(pldv)
$$
$$
t_\pm^0=-{1 \over4 (x_\pm)^2} \eqno(tpm)
$$
corresponds to
$$
\Om=-\la^2x^+x^--{1\over4}\ln[-4\la^2x^+x^-]. \eqno(oldv)
$$
The solution corresponding to general incoming matter from
$\cI^-$ is
$$
\eqalign{
\Om &=-{\la^2}x^+(x^-+{1\over\la^2}P_+(x^+))+{1\over\la}M(x^+)\cr
&-{1\over4}\ln[-4\la^2x^+x^-], \cr} \eqno(gsol)
$$
where
$$
\eqalignno{
M(x^+) &= \la\int_0^{x^+} d \tilde x^+ \tilde x^+(T^M_{++}-t^0_+), &(mdef)\cr
P_+(x^+) &=\int_0^{x^+} d \tilde x^+(T^M_{++}-t^0_+). &(pdef) \cr}
$$
and $t_-=t_-^0$.
By transforming back to $\rho,\ \phi$ variables it can be seen for large $M$
that this corresponds at early times to a black hole which forms and
evaporates.

However, the late-time behavior of \(gsol) is unphysical. Viewed as a function
of $\phi$, $\Om$ has a minimum at
$$
\eqalign{
\phi_{cr} &=-{1\over2}\ln{N\over 48},\cr
\Om_{cr} &={1\over 4}. \cr}
$$
There is no real value of $\phi$ corresponding to $\Om<\Om_{cr}$.
$\Om=\Om_{cr}$
should be regarded as the analog of the origin of radial coordinates
and the end
of the spacetime, rather
than continuing to negative radius. Reflecting boundary conditions,
consistent with energy conservation should
be imposed. RST accordingly  require
$$
\eqalign{
f_i&|_{\Om=\Om_{cr}} =0,\cr
\part_\pm\Om&|_{\Om=\Om_{cr}}=0.\cr}\eqno(rbc)
$$
The line $\Om=\Om_{cr}$ along which the boundary conditions are imposed
undergoes dynamical motion in
the $x^+,x^-$ plane. Of course this boundary  line
could be moved to a fixed timelike coordinate line {\it e.g.\/} $x^+=x^-$ by a
conformal transformation. However, this would be incompatible with Kruskal
gauge and does not simplify the analysis.

It follows from the equations of motion that the boundary curve $\hat x^-(x^+)$
obeys
$$
\la^2\part_+\hat x^-(x^+)=-\part_+P_+(x^+)+{1\over4(x^+)^2}.\eqno(ccc)
$$
If $\part_+P_+$ is small enough, the right hand side is positive and the
boundary
curve is a timelike line. No black holes are formed: incoming matter is
benignly reflected up to future null infinity. A similar behavior occurs in
four-dimensional general relativity in that sufficiently weak scalar $S$-waves
can simply pass through  the origin without collapse.

On the other hand, if $\part_+P_+$ exceeds the critical value $1/4(x^+)^2$, the
boundary curve turns to the right (towards spatial infinity) and becomes
spacelike as in the shock wave geometry of figure 1. It can be seen
that the scalar curvature diverges along the spacelike
segments of the boundary curve. It is not possible to
implement the boundary condition
\(rbc) along these segments. Such spacelike boundary segments
necessarily bound regions of future trapped points
where $\part_+\Om<$ and $\part_-\Om<0$, which is
the interior of a black hole. Thus these spacelike singularities
resemble in every way the singularities inside four-dimensional black holes. We
regard the singularity as a boundary of the spacetime, and information which
flows
into this  spacelike  curve is lost to asymptotic observers\footnote{*}{We are
essentially defining - as opposed to deriving - the model to lose information
by this boundary
condition. Physics at the singularity is of course not controlled by
our approximations, and other
boundary conditions might be possible. For example one might imagine a
(causality violating) conveyor belt which moves the information along the
singularity and then releases it at the endpoint. It would be of great interest
to find alternate boundary conditions which consistently implement this idea.}.

The trajectory of a spacelike segment of the boundary curve is determined, not
by boundary conditions, but by the initial conditions on $\cI^-$. If the
incoming energy is finite, the boundary curve will eventually revert to a
timelike  trajectory. This is the ``endpoint" at which the future apparent
horizon---the boundary dividing the regions $\part_+\Om>0$ and
$\part_+\Om<0$---meets the singularity, and the black hole has evaporated to
zero size. After the endpoint the boundary  conditions \(rbc) are immediately
imposed[\cite{rst, AS}]. This is depicted in figure~1.

These boundary conditions have an important consequence {\it at} the
endpoint\footnote{$\dagger$}{RST seem to
inexplicably alter their boundary
condition at the endpoint.}, as follows.
Parameterizing the boundary curve as  $\hat
x^+(x^-)$, the constraints imply
$$
\la^2\part_-\hat x^+=-\part_-P_-+{1\over4(x^-)^2}. \eqno(bcm)
$$
Following the boundary curve down from above the endpoint, we see that it makes
a sudden turn to the right (away from spatial infinity) at the endpoint. This
is consistent with \(bcm) only if $\part_-P_-$, or equivalently
$T^M_{--}$ is negative and infinite at the
endpoint. RST indeed find in the exact solution that $T^M_{--}$ must have a
negative delta-function. Since the
$f_i$ are zero on the boundary, this implies that
$t_-$ has a delta-function at the endpoint value of $x^-$.

A general lesson, which will be important in the following, can be extracted
from the preceding discussion. Throwing positive energy in from $\cI^-$
causes the boundary curve to lean towards spatial infinity. Above some
threshold, it actually causes it to become spacelike. Throwing negative
energy at the boundary from $\cI^-$ causes it to recede. Since it is
receding, throwing more negative energy at the boundary has less effect. It
can be forced to make a spacelike turn to the left only by an infinitely
negative energy density.

Returning to the discussion of the endpoint, a delta function in $T^M_{--}$
corresponds to a signal which travels from the endpoint up to $\cI^+$. This
signal is referred to as a ``thunderpop".
The thunderpop is  an essentially planckian object, since it involves
high frequencies and emerges from a high-curvature region.
Ordinarily in a $1/N$ expansion wavelengths are restricted to
be greater then $1/N$, since at shorter distances energy fluctuations
become large and the $1/N$ expansion breaks down. Thus thunderpop
type states would not usually be included in the Hilbert space.
However here we see that they generically arise from the evolution of
long-wavelength states, and so must be considered. This is a breakdown
of decoupling. Of course as the detailed structure of the thunderpop
is sensitive to short distance physics, the description we give here
is approximate. At the very least the thunderpop should be smeared over
a region of size $1/N$.

 Since $ f_i =0$,
yet $T^M_{--}$ (which should be understood as an expectation value)
is non-zero and negative
along the thunderpop, it is a ``highly quantum'' state of $f$ - modes.
Negative $T^M_{--}$ can arise from
off-diagonal terms in the expectation value [\cite{prtr}].
However there are restrictions on how negative $T^M_{--}$
can be: the total energy must be positive for reasonable states,
and a region of negative $T^M_{--}$ can not be ``too far'' from a
region of positive  $T^M_{--}$. Thus it is not obvious that the
thunderpop can really be represented as a semiclassical quantum state.
In the appendix we demonstrate that it is nevertheless possible.

It is important to keep in mind that, although the RST model has an actual
spacelike singularity, this will not be important in the following because the
$\slS$-matrix is insensitive to all physics behind the global horizons.
One can certainly imagine
altering the RST model in such a way so as to
eliminate the spacelike singularity and store the information forever inside
the black hole, or allow it to be carried away by a baby universe. The
following discussion would apply equally to such models.

In conclusion, the RST model embodies all the features of black hole
evaporation anticipated by Hawking. Black holes form and evaporate in a finite
time, leaving nothing behind. Information is lost behind a global event
horizon.
The only unanticipated feature is the generic occurrence of a thunderpop at the
evaporation endpoint. In section~4 we shall see this plays a key role in the
restoration  of CPT invariance.

\head{3. The RST $\slS$-Matrix}

Given the quantum state on $\cI^+$, it is not in general possible to
reconstruct the quantum state on $\cI^-$. Additional information, corresponding
to the
state swallowed by the black hole, is required. One therefore expects that
only a $\slS$-matrix, rather than an $S$-matrix, exists for the RST model. It
is perhaps possible to obtain analytic formulae for the matrix elements to
leading order in large-$N$ using technology developed in
[\cite{gine}]\footnote{$\dagger$}{This has been partially worked out
for shock wave geometries by L. Thorlacius (private communication).} .
However, such explicit formulae will
not be required for our considerations, and we
will content ourselves in this section with a general
description of the large-$N$ $\slS$-matrix.

Incoming states are created by acting with $f$-oscillators
on the $f$-vacuum.
%\footnote{*}{Recall
%that in the gauge \(gchc)\ the flat line element is
%$ds^2=-dx^+dx^-/x^+x^-$. \(fosc)\ is valid in the gauge \(gchc),
%whichis the analog of light cone gauge in string theory. The
%operator $\Om$ can be reexpressed in terms of the physical
%degrees of freedom $f_i$ by using
%the constraints \(cstr), with integration constants fixed by \(oeom).
%Expectation values of operators will then
%obey all equations of motion and constraints. }
%$$
%a^+_{iw}=\int^\infty_0dx^+(x^+)^{iw-1}f_i(x^+) \eqno(fosc)
%$$
%$f_i$ here is, unlike in the previous section, understood to be an
%operator.
Let us first consider states obtained by acting on the
vacuum with order $N^0$ creation operators. It follows from \(nact) or \(tdef)
that there is no back-reaction on the metric or dilaton to leading order in
$1/N$. This is simply because ``Newton's constant" is taken to zero as $1/N$.
Thus the scattering is simply obtained by reflecting the $f$-particles off of
the boundary at $\Om=\Om_{cr}$, (according to \(rbc)), which in the vacuum is
the timelike  line $4\la^2x^+x^-=-1$.
Clearly a unitary $S$-matrix exists in this sector of the Hilbert space.

In order to investigate black hole dynamics, one must consider states with
incoming energies (\ie $\langle T^M_{++} \rangle$)
of order one. Semiclassical configurations corresponding to
collapsing matter
are typically described by
specifying $c$-number initial data $\rho^c(x^+),~~$$\phi^c(x^+),~~$$f^c_i(x^+)$
and
$t_+(x^+)$ on $\cI^-$. This description
is redundant because of conformal gauge
invariance. The semiclassical quantum state  corresponding to
this  initial data has the simplest
description in the gauge $t_+=0$ (the gauge transformation
properties of $t_+$ are described in the appendix).  In this gauge it can
be represented on $\cI^-$ as a coherent state of the form:
$$
|f_i^c\rangle=A:e^{-{i \over 2 \pi}\sum^N_{i=1}\int dx^+\part_+f_i^c(x^+)
 f_i(x^+) }:|0\rangle ,
\eqno(chst)
$$
where the normal ordering and the vacuum are defined with respect to
$t_+=0$ coordinates and $A$ is a normalization factor.  Its time evolution is
then given - to
leading order in $N$ - by classical evolution of $f^c_i$.
In general the coherent states are an overcomplete basis, but in the
semiclassical large-$N$ limit $A$ becomes small and
they become orthogonal.

In
$t_+=0$ coordinates the Kruskal condition
\(gchc) will of course not hold in general. Rather one has
$$
\Om=\chi +\om , \eqno(tgc)
$$
where $\om$ is a $c$-number solution of the free wave equation. In this
$t_+=0$ gauge the
operator constraints become
$$
-\part_+^2\om- \part_+\om \part_+\om +2\part_+\om \part_+\Om +\part^2_+\Om =
-{6\over N}\sum^N_{i=1}\part_+ f_i\part_+ f_i . \eqno(tcs)
$$
$\Om$ is then found in terms of the physical degrees of freedom
$f_i$ by integrating this equation, with some integration
constants fixed by \(oeom)\footnote{*}{This is the analog of light-cone
gauge in string theory. A DDF-like description, in which
$\chi$ and $\Omega$ are independent operators, is also
possible[\cite{verl,deal}].}.

If the $c$-number field configurations $f_i^c$ in \(chst) are such that
$$
\langle T^M_{++} \rangle={6\over N}\sum_{i=1}^N\part_+f_i^c\part_+f^c_i
\eqno(texp)
$$
is order $N^0$, there is a
leading order back reaction on the geometry, and $\rho$ and $\phi$ are
non-zero. If $\rho$ is non-zero, then the representation \(chst)\ of the
incoming state will not correspond to the natural Fock basis of
particles which might be seen by asymptotic inertial observers.
The Bogoliubov transformation associated to the coordinate transformation
from the Kruskal gauge \(gchc)\ to $\rho=0$ gauge can be used to
reexpress the state $|f_i^c\rangle$ in this basis. It will not in general be a
simple coherent state in this asymptotic Fock basis.

If $\langle T^M_{++} \rangle$ is below threshold for black hole production, the
semiclassical
geometry is obtained by solving the equations \(oeom) with initial conditions
determined by the expectation value of
\(tcs) and boundary conditions of the form \(rbc)
suitably modified to conform to the gauge \(tgc). The outgoing
coherent state
on $I^+$ into which $ |f_i^c\rangle$
scatters is found by reflection off of this
dynamically determined boundary. This will naturally lead to an expression in
terms of oscillators which are positive frequency with
respect to $t_+=0$ coordinates on $\cI^+$. However, in
the process of solving for the semiclassical geometry, the conformal factor
$\rho$ is altered on $\cI^+$. Thus a
Bogoliubov transformation is required to obtain the outgoing state in a natural
Fock basis on $\cI^+$.

Let us now consider further acting on the state $|f_i^c\rangle$ in \(chst) with
order $N^0$ $f$ creation operators. To leading order in $1/N$, this will not
affect the stress tensor \(texp) and the semiclassical geometry and boundary
curve will be unaltered\footnote{*}{Energy conservation should be restored by
including the back reaction at the next order in $1/N$.}.
The outgoing state is  then determined by boundary
reflection and Bogoliubov transformation. In this manner the $S$-matrix for
states obtained by acting with order $N^0$ oscillators on \(chst) can be
constructed.

If $\langle T^M_{++} \rangle$ exceeds the critical threshold, then the boundary
curve will have
a singular spacelike segment. However,  the incoming state on $\cI^-$ will
still scatter (given the endpoint prescription of section 2) to a definite
out-state on $\cI^+\cup\cH^+$,
where $\cH^+$ is the future event horizon.
States obtained by acting on the incoming coherent state with order $N^0$
$f$-oscillators will scatter to a definite pure out-state on
$\cI^+\cup\cH^+$. These out-states are of the general form
$$
|\Psi_{out}\rangle=\sum_{x, \a}\Psi_{x\a}|x\rangle|\a\rangle \eqno(sdec)
$$
where $|x\rangle(|\a\rangle)$ is a state on $\cI^+~(\cH^+)$. The ``$S$-matrix"
for
such states  has the form
$$
\langle x|\langle\a|S|a\rangle={S_{x\a}}^a. \eqno(smat)
$$
In this expression, $|a\rangle$ denotes an incoming state which collapses to
form a black  hole, $|\a\rangle$ denotes a state on $\cH^+$, and $|x\rangle$
denotes a state on $\cI^+$ which results from black hole
evaporation---including a thunderpop.

The RST dollar matrix can now be constructed by tracing over states on $\cH^+$
$$
{{{{{(\slS_{rst})}_x}^y}_b}^a}\equiv {S^{\dagger y\a}}_b
{S_{x\a}}^a. \eqno(drst)
$$
$\slS_{rst}$ conserves probability
$$
{{{{{(\slS_{rst})}_x}^x}_b}^a}={{\de_b}^a}. \eqno(pcon)
$$
However, it is not the product of two unitary matrices, and pure states will
evolve into mixed ones.

\head{4.  CPT Invariance}

It is easy to see that $\slS_{rst}$ is not, on its own, weakly CPT invariant.
The CPT transform of the incoming states which collapse to form a  single
black hole
will never appear  as an out-state on $\cI^+$:
$$
\langle{x|\Th|a}\rangle=0. \eqno(hrth)
$$
This is because the out-states all contain planckian thunderpops where
$\langle{T^M_{--}}\rangle$ diverges. Also $\langle{T^M_{++}}\rangle$ always
exceeds the critical value
somewhere on $\cI^-$ (otherwise a black hole would not form),
but $\langle{T^M_{--}}\rangle$ will generically not
exceed the critical value on $\cI^+$ (otherwise a white hole would be in the
past of $\cI^+$). The orthogonality condition \(hrth) between in
and out Hilbert
spaces immediately precludes the possibility of weak
CPT invariance[\cite{wald}].

Stated in this manner, the remedy is obvious. One must include in  the incoming
Hilbert space thunderpop states which create white holes, as
illustrated in figure 2. One may
then construct the large-$N$ $\slS$-matrix
$$
\slS_{cpt}=\slS_{rst}+\Th\slS^T_{rst}\Th. \eqno(sccpt)
$$
The first term acts on states which collapse  to form black holes, while the
second acts on thunderpop states which create  white holes. It is easy to
check,
in bases for which $\Th^{-1}=\Th^T$, that $\slS_{cpt}$ obeys the weak CPT
condition \(wcpt) as well as probability conservation \(pcon).

Physically, the second term in \(sccpt) can be viewed as follows. A thunderpop
carrying negative, divergent $\langle T^M_{++} \rangle$ arrives at the boundary
curve,
forcing it to turn left (away from spatial infinity) and become spacelike. The
future of this point will then depend on boundary conditions along the
null surface $\cH^-$. One must sum over all possible states on $\cH^-$, and
construct  a $\slS$-matrix by tracing over these states. The measure for this
trace is fixed by weak CPT invariance.

Note that the existence of the thunderpop appears crucial to the
restoration of CPT invariance. If the black holes
just quietly disappeared into the
vacuum, it is hard to see how the CPT reverse of the outgoing state
could form a white hole. The thunderpop is required to nucleate the
white hole into which the CPT-reversed Hawking radiation subsequently falls.
This makes it difficult to see how CPT could be realized in a long-wavelength
effective theory in which thunderpops are integrated out.
It would certainly be interesting to understand the
analog of thunderpops in four dimensions. CPT would seem to require
that there is something which can nucleate a white hole.
Perhaps planckian negative energy densities are necessary for
four as well as two-dimensional white hole nucleation.

The $\slS$-matrix is naturally represented in the language of functional
integrals. The scattering amplitude from a coherent state $|f(\cI^-)\rangle$
 on $\cI^-$ which collapses to form a black
hole to an out-state $|f''(\cH^+),f'(\cI^+)\rangle$ is given by
$$
S(f''(\cH^+),f'(\cI^+);f(\cI^-))=
\int^{f'(\cI^+),f''(\cH^+)}_{f(\cI^-)}\cD f e^{-iS}, \eqno(sfnt)
$$
where the boundary conditions are those appropriate to coherent states.
The primes on the boundary data $f$ indicate that they are generally different
functions. The matrix element $\slS_{rst}$ between an initial coherent density
matrix
$|f\rangle\langle\tilde f|$ and a final
coherent density matrix $|\tilde f'
\rangle \langle f'|$is then obtained by multiplying \(sfnt) by its complex
conjugate, and functionally integrating over $f''(\cH^+)$ ({\it i.e.,\/}
tracing
over states on the horizon)
$$
\eqalign{\slS[\tilde f'(\tilde \cI^+), &f'(\cI^+); \tilde f
(\tilde\cI^-), f(\cI^-)]\cr
&=\int \cD f''
\int^{f'(\cI^+),f''(\cH^+)}_{f(\cI^-)}\cD f e^{-iS}
\int^{\tilde f'(\tilde\cI^+), f''(\tilde\cH^+)}_{\tilde
f(\tilde\cI^-)}\cD \tilde f e^{iS}.\cr}
\eqno(ssfnt)
$$
The integration over common values of boundary data
$f''$ on $\cH^+$ and $\tilde\cH^+$ sews together the two path
integrals along the horizons. This may be represented as a single
functional integral on the
unusual semiclassical geometry depicted in figure~3
(corresponding to white and black hole spacetimes sewn together at the
horizons):
$$
\slS[\tilde f'(\tilde \cI^+), f'(\cI^+); \tilde f
(\tilde\cI^-), f(\cI^-)]
=
\int^{f'(\cI^+),\tilde f'(\tilde\cI^-)}_{f(\cI^-),\tilde
f(\tilde\cI^+) }\cD  fe^{-iS}.\eqno(ggg)
$$
It should be understood in \(ggg) that the time flows forward
from $\tilde\cI^+$ to $\tilde\cI^-$ and the sign of the
action is reversed in the double of
the original spacetime .

Let us summarize the rule for computing the large-$N$ dollar
matrix element $\slS[\tilde f'(\tilde \cI^+), f'(\cI^+); \tilde f
(\tilde\cI^-), f(\cI^-)]$ for the case in which $|f\rangle$ and
$|\tilde f\rangle$ both collapse to form a black hole, and $|f'\rangle$ and
$|\tilde f'\rangle$ are both outgoing thunderpop states.
The first step is to determine the semiclassical geometry. To do so, construct
the stress tensor corresponding to $f(\cI^-)$, and determine the incoming
geometry by solving the constraints. Then solve the semiclassical
equations of motion with this initial data. This will determine the full
semiclassical spacetime geometry and field configurations, including
$f''(\cH^+)$ and $f'(\cI^+)$. The outgoing state is (in $t_+=0$ gauge)
the coherent state associated to this semiclassical data.
The large-$N$ approximation to the $\slS$-matrix
vanishes unless the semiclassical
field configuration $f'(\cI^+)$ so obtained agrees with its argument
because coherent states constructed
with different $c$-number functions are orthogonal at large $N$. Next,
solve the constraints using  $\tilde f'(\tilde\cI^+)$ to get (CPT reversed)
initial data for $\tilde\cI^+$. This initial data
will produce a white hole. Use the CPT reverse of the final data $f''(\cH^+)$
obtained previously on $\cH^+$ as initial data on the white hole horizon on
$\tilde\cH^+$. One may then compute $\tilde f(\tilde\cI^-)$ as the CPT
reverse of the  final data in this spacetime. The semiclassical coherent state
$\slS$-matrix element
vanishes if the semiclassical data thereby obtained does not agree with its
argument.

$\slS$-matrix elements of states obtained by acting on these
coherent states with order $N^0$ oscillators are now simply $S$-matrix elements
(remembering
 to perform appropriate Bogoliubov transformations) in
the sewn spacetime of figure 3.

So
far we have discussed processes involving one black hole or one white hole. The
construction of the $\slS$-matrix for two black holes or the CPT reverse, two
white holes proceeds similarly. Now there are two spacelike segments along
which the two spacetimes must be sewn together.

Clustering however requires that we include initial states with both black
holes and white holes. This can result, after sewing, in a spacetime with
closed timelike loops as in figure~4. New features, which we have not yet
analyzed, may arise in computing the $\slS$-matrix from $S$-matrix elements on
such spacetimes.

This prescription for computing the $\slS$-matrix formally generalizes to
four dimensions: Form the double of a  spacetime by sewing its black (white)
hole horizons to the white (black) hole horizons of its
CPT conjugate. The $\slS$-matrix is then obtained from the $S$-matrix
on the doubled spacetime. This picture suggest a
natural way of including the effects of virtual black holes:
they are simply wormholes which connect the
original spacetime to its double. For the usual reasons[\cite{worm}],
this will {\it not} lead to quantum incoherence, but rather will
turn coupling constants into dynamical variables. The only slight difference is
that the wormhole $\a$-parameters label
$\slS$-matrix rather than $S$-matrix superselection sectors. This leads us
to question the assertion[\cite{hwktwo, bps}] that information
loss from real black holes inevitably implies that virtual black holes lead
to tiny unitarity violations in a long distance effective theory.

\head{5. Some Problems}

In this paper we have described the black and white hole
superscattering matrix to leading order in a $1/N$ expansion.
Going beyond leading order may not be straightforward for several
reasons.

The first is that the RST boundary condition - which entered crucially
into our discussion - has only been fully defined at the semiclassical level.
Heuristically one might hope to define the full quantum theory by
restricting the range of functional integration of the field
$\Om$. However it is not clear what this means in practice, or if the
resulting theory will conserve energy. Furthermore virtual fluctuations
of $\Om$ about its vacuum value will lead to internal boundaries, or
holes in spacetimes. These difficulties will certainly arise at the
non-perturbative level, and perhaps even in $1/N$ perturbation theory.

A second problem is the white hole instabilities
discussed in [\cite{wbld}]. This problem has also been stressed
- in a slightly different form - in more recent papers[\cite{brt,verl}].
Consider a particular white hole formation/evaporation process obtained as
the time reverse of black hole formation/evaporation, projected on
to  some particular
pure state on $\cH^-$ and $\cI^-$ . Now consider adding a single
incoming $f$ quanta to the state on $\cI^-$ somewhere near or shortly after
the incoming thunderpop. It is straightforward to show that the energy
(\ie $T^M_{++}$) of this
state is blueshifted by an amount of order $e^{M/\la}$,
where $M$ is the white hole mass.
The energy of the quanta is then of order $e^{M/\la}/N$.

In the large-$N$ limit the back reaction of this quanta can be neglected.
However if $N$ is large but finite {\it and} $M$ is greater than $\ln N$,
the back reaction is not small.
The $1/N$ expansion presumably breaks down for such matrix elements
at such large values of $M$.

Physically one expects, for sufficiently large $M$, that the
blueshifted quanta will lead to a black hole after the white hole.
Thus beyond the $1/N$ expansion one does not expect in general
to always be able to determine the numbers of
black and white holes from the state on $\cI^-$
alone: knowledge of the state on $\cH^-$ is also required.
This will bring new features into the analysis, and we do
not know how one develops a systematic perturbation expansion in
this situation.

In conclusion, the inclusion of white holes restores CPT
to leading order in a $1/N$ expansion for processes with one white hole or one
black hole. The general case remains unresolved. We hope to have convinced the
reader that it is an interesting problem.

\head{Acknowledgements}

The idea of investigating CPT in the RST model arose in discussions with
S. Hawking, to whom I am grateful for numerous conversations, and
who has obtained related results. I
also benefited from discussions with  S. deAlwis, S. Giddings, J. Polchinski,
J. Preskill, L. Thorlacius, S. Trivedi, H. Verlinde and R. Wald.
This work was supported in part by DOE grant DOE-91ER40618 and NSF grant
PHY-89-04035.

\head{Appendix}

In this appendix we will describe the RST
thunderpop [\cite{rst}] as a quantum state of
$f$-particles.

For an incoming shock wave ${\la (T^M_{++}-t^0_+)}=M\delta(x^+-1)$ ,
the field $\Om$ in Kruskal gauge is given by
$$
\eqalign{
\Om &=-{\la^2}x^+(x^-+{M\over\la^3})+{M\over\la}\cr
&-{1\over4}\ln[-4\la^2x^+x^-], \cr} \eqno(gsolp)
$$
in the region above the shock wave at $x^+=1$, where $\Om >\Om_{cr}$
and prior to the thunderpop. The thunderpop orignates at the
black hole endpoint
$$
\Bigg( x^-_s,x^+_s\Bigg) =\Bigg( -{M \over \la^3(1-e^{-4M/\la})},
{\la \over 4M}(e^{4M/\la}-1)\Bigg),\eqno(xs)
$$
where the apparent horizon and singularity meet. For $x^- >x^-_s$,
the solution is joined onto the shifted linear dilaton vacuum
$$
\Om =-{\la^2}x^+(x^-+{M\over\la^3})
-{1\over4}\ln[-4\la^2x^+(x^-+{M\over\la^3})].  \eqno(glp)
$$
$\Om$ is continuous along the thunderpop at $x^- = x^-_s$.
\(glp) and \(gsolp) obey the constraints \(cstr) with
$$
\eqalign{t_-&=-{1 \over 4
{x^-}^2}~~~~~~~~~~~~~~~~~~~~~~~~~~~~~~~~~~~~~~~~~~~~~~~x^- < x^-_s,\cr
          t_-&=-{\la^3 e^{4M/\la} \over 4M} (1-e^{-4M/\la})^2
         \delta(x^- - x^-_s)~~~~~~~~~x^- = x^-_s,\cr
    t_-&=-{1 \over 4 (x^-
+{M\over\la^3})^2}~~~~~~~~~~~~~~~~~~~~~~~~~~~~~~~~~~~~~~x^- > x^-_s.\cr}
\eqno(zzz)
$$

The metric corresponding to \(glp) and \(gsolp) is asymptotically flat on
$\cI^+$. Quantum states on $\cI^+$ are best
described in terms of creation and annihilation operators defined
with respect to the natural flat coordinates.
Defining
$$
\eqalign{
y^-&=-\ln \Bigg[ { \la^3 x^-+M \over \la^3 x^-_s+M }\Bigg] ,\cr y^+&=\ln (\la
x^+),\cr}\eqno(xxyy)
$$
one finds that asymptotically on $\cI^+$
$$
ds^2 \rightarrow -dy^+dy^-. \eqno(syy)
$$
In these coordinates, the radiation flux on $\cI^+$ is
given by $t_-$. To determine $t_-$, note that the constraints
in a general gauge are
$$
2(e^{-2\phi}+{1 \over 4})\nabla^2_- \phi
={1 \over 2}\sum^N_{i=1}\part_- f_i\part_- f_i+
{N \over 12}(t_- -\part_-\rho \part_- \rho +\part_-^2 \rho ),\eqno(gcst)
$$
where $\nabla_-$ is the covariant derivative. Both sides of this equation
must transform covariantly under a conformal coordinate transformation
$\tilde x^-= \tilde x^- (x^-)$. This requires that
$$
(\part_-\tilde x^-)^2 \tilde t_-= t_- -(\part_- \tilde x^-)^{1/2} \part_-^2
(\part_- \tilde x^-)^{-1/2}. \eqno(ttr)
$$
One then finds that in $y$ coordinates
$$
\eqalign{t_-&={2 e^{y^-}(e^{4M/\la}-1)+
e^{2y^-}(e^{4M/\la}-1)^2 \over 4(1+e^{y^-}(e^{4M/\la}-1))^2}
{}~~~~~~~~{y^-}<0,\cr
          t_-&=-{1 \over 4} (1-e^{-4M/\la})
         \delta({y^-})~~~~~~~~~~~~~~~~~~~~~~~~~~~{y^-} = 0,\cr
    t_-&=0~~~~~~~~~~~~~~~~~~~~~~~~~~~~~~~~~~~~~~~~~~~~~~~~~~~~~~~{y^-} > 0.\cr}
\eqno(yyy)
$$

A thunderpop state $|T\rangle$
clearly can not be described as a coherent state in the Fock basis
associated to $y$ coordinates because the energy density of such
states is non-zero only when $\langle f_i \rangle$ is non-zero.
To represent $|T\rangle$ as a coherent state, we must find a coordinate
system in which $t_-$ vanishes. This is accomplished by
solving \(ttr) with the left hand side set to zero and
$t_-$ given by \(yyy). One finds that $t_-$ vanishes in
$z$ coordinates
$$
\eqalign
{z^+&=y^+,\cr
 z^-&=-\ln\Bigg[ e^{-{y^-}-4M/\la}+1-e^{-4M/\la} \Bigg]~~~~~{y^-}<0,\cr
 z^-&={4 {y^-}e^{-4M/\la} \over 4+(1-e^{-4M/\la}){y^-}}~~~~~~~~
{}~~~~~~~~~~~~~{y^-}>0. \cr}
\eqno(zzyy)
$$
A thunderpop may then be represented as the $z$ vacuum
$$
|T\rangle =|0_z\rangle,\eqno(tpop)
$$
where $|0_z\rangle$ is annihilated by modes of the $f$ field which are
negative frequency with respect to $z^-$. To represent
this state in the $y$ Fock basis - which corresponds to
particle states detected by inertial asymptotic observers - one must
perform the Bogoliubov transformation corresponding to \(zzyy).
A nice discussion of these transformations in two dimensions
can be found in [\cite{gine}].

Note that an isolated thunderpop - unaccompanied by nearby
positive energy density - can not be represented by a quantum state
as in \(tpop). The analog of the $z$ coordinates in which $t_-$
vanishes would not cover all of $\cI^+$. Thus the global
existence of $z$ coordinates is an important consistency check
for the RST model.

Finally, we would like to remind the reader that the state \(tpop)
is not the full outcome of shock wave collapse. The state on
$\cI^+$ will be a mixed state, or density matrix. To find that
density matrix one must first compute the Bogoliubov transformation from the
incoming pure state on $\cI^-$ to the outgoing pure state on
$\cH^+ \cup \cI^+$, and then trace over states on $\cH^+$.
\(tpop) is presumably one of the pure states which
appears on $\cI^+$ with some probability.

\head{References}

\refis{CGHS} C. G. Callan, S. B. Giddings, J. A. Harvey and A. Strominger,
 {\sl Phys.\ Rev. D  } {\bf 45}
(1992) R1005; For recent reviews see J. A.
Harvey and A. Strominger,
{\it Quantum Aspects of Black Holes} preprint EFI-92-41, hep-th/9209055, to
appear in the proceedings of the 1992 TASI Summer School in Boulder, Colorado,
and S.B. Giddings, {\it Toy Models for Black Hole Evaporation} preprint
UCSBTH-92-36, hep-th/9209113, to appear in the proceedings of the
International Workshop of Theoretical Physics, 6th Session, June 1992, Erice,
Italy.

\refis{page}D. N. Page,  {\sl Phys.\ Rev. Lett.} {\bf 44}
(1980) 301.

\refis{bos} T. Banks, M. O'Loughlin and A. Strominger, {\sl Phys.\ Rev. D  }
{\bf 47}
(1993) 4476.

\refis{emod} A. Bilal and C. G. Callan, {\sl Nucl. Phys. B} {\bf 394}
(1993) 73; S.P. deAlwis,
 {\sl Phys. Lett. B} {\bf 289\/} (1992) 278;
 {\sl Phys. Lett. B} {\bf 300\/} (1993) 330;
 S. B. Giddings and A. Strominger, {\sl Phys. Rev. D} {\bf 46\/} (1993) 2454.

\refis{AS} A. Strominger, unpublished.

\refis{rst} J. G. Russo, L. Susskind, and L. Thorlacius, {\sl Phys. Rev.
D\/}{\bf 46\/} (1992) 3444; {\sl Phys. Rev. D\/} {\bf 47\/} (1993) 533.

\refis{sork} R. M. Sorkin, unpublished.

\refis{prtr}J. Preskill and S. Trivedi, to appear.

\refis{gine} S. B. Giddings and W. Nelson,
{\sl Phys. Rev. D} {\bf 46\/} (1992) 2486.

\refis{lowe} D. A. Lowe,
{\sl Phys. Rev. D\/} {\bf 47} (1993) 2446.

\refis{hawk} S.~W.~Hawking,
 {\sl Comm. Math. Phys.} {\bf 43} (1975) 199;
 {\sl Phys. Rev. D} {\bf 14} (1976) 2460.

\refis{verl}E.~Verlinde and H.~Verlinde,
{\it A Unitary S-matrix for 2D Black Hole Formation and Evaporation,}
Princeton preprint, PUPT-1380, IASSNS-HEP-93/8;
 K. Schoutens, E.~Verlinde, and H.~Verlinde,
{\it Quantum Black Hole Evaporation,} Princeton preprint,
PUPT-1395, IASSNS-HEP-93/25.

\refis{pira}T.~Piran and A.~Strominger, {\it Numerical Analysis of Black Hole
Evaporation,} hep-th/9304148, ITP preprint NSF-ITP-93-36.

\refis{deal}S. P. deAlwis, to appear.

\refis{hwktwo}S. W. Hawking,
{\sl Comm. Math. Phys.} {\bf 87} (1982) 395.

\refis{worm}S. Coleman {\sl Nucl. Phys. B } {\bf 307} (1988) 867;
S. B. Giddings and A. Strominger  {\sl Nucl. Phys. B } {\bf 307} (1988) 854.

\refis{bps} T. Banks, M. E. Peskin and L. Susskind,
{\sl Nucl. Phys. B } {\bf 244} (1984) 125.

\refis{wald} R. M. Wald, {\sl Phys. Rev. D} {\bf 21} (1980) 2742.

\refis{wbld} S. Ramaswamy and R. M. Wald, {\sl Phys. Rev. D}
{\bf 21} (1980) 2736.

\refis{brt}G. `t Hooft, {\sl Nucl. Phys. B } {\bf 335} (1990)
138 and references therein;
T. Jacobson, {\sl Phys.\ Rev.  D\ }44 (1991) 173; {\it
Black Hole Radiation in the
Presence of a Short-Distance Cutoff} UMDGR93-32, ITP preprint (1993).

%\refis\past{Y.~Park and A.~Strominger
% Phys. Rev. & D47 (93) 1569.}

\endreferences

\endit